\documentclass{aa}  
\usepackage{graphicx}
\usepackage{txfonts,natbib}
\bibpunct{(}{)}{;}{a}{}{,}
\begin{document}
   \title{New algorithms for adaptive optics point-spread function reconstruction}

   \author{E. Gendron
          \inst{1}
          \and
          Y. Cl\'enet\inst{1}
          \and
          T. Fusco\inst{2}
          \and
          G. Rousset\inst{1}
          }
\titlerunning{New algorithms for AO PSF reconstruction}

   \offprints{Y. Cl\'enet}

   \institute{Observatoire de Paris, LESIA, 5 place Jules Janssen, F-92195 Meudon Cedex, France\\
              \email{eric.gendron@obspm.fr, yann.clenet@obspm.fr, gerard.rousset@obspm.fr}
              \and
   ONERA, BP 52, 29 avenue de la Division Leclerc, 92320 Ch\^{a}tillon Cedex, France\\
              \email{thierry.fusco@onera.fr}
             }

   \date{Received 6 March 2006; Accepted 20 June 2006}

 
  \abstract
   {The knowledge of the point-spread function compensated by adaptive optics is of prime importance in several image restoration techniques such as deconvolution and astrometric/photometric algorithms. Wavefront-related data from the adaptive optics real-time computer can be used to accurately estimate the point-spread function in adaptive optics observations. The only point-spread function reconstruction algorithm implemented on astronomical adaptive optics system makes use of particular functions, named $U_{ij}$. These $U_{ij}$ functions are derived from the mirror modes, and their number is proportional to the square number of these mirror modes.}
   {We present here two new algorithms for point-spread function reconstruction that aim at suppressing the use of these  $U_{ij}$ functions to avoid the storage of a large amount of data and to shorten the computation time of this PSF reconstruction. }
   {Both algorithms take advantage of the eigen decomposition of the residual parallel phase covariance matrix. In the first algorithm, the use of a basis in which the latter matrix is diagonal reduces the number of $U_{ij}$ functions to the number of mirror modes. In the second algorithm, this eigen decomposition is used to compute phase screens that follow the same statistics as the residual parallel phase covariance matrix, and thus suppress the need for these $U_{ij}$ functions.}
   {Our algorithms dramatically reduce the number of $U_{ij}$ functions to be computed for the point-spread function reconstruction. Adaptive optics simulations show the good accuracy of both algorithms to reconstruct the point-spread function.}
   {}


   \maketitle
%

\section{Introduction}

Since the advent of adaptive optics (AO), it has been possible to perform imaging with a spatial resolution very close to the diffraction limit. Despite this large improvement, the correction of the atmospheric turbulence by an AO system is only partial, and point source images are for example still affected by a halo that surrounds them.

Such effects of the partial AO correction can be corrected by image restoration techniques such as deconvolution algorithms. Except in the "blind" deconvolution case \citep{fusco99,christou99}, these algorithms need an estimation of the point-spread function (PSF), or its corresponding optical transfer function (OTF) derived from the PSF by a Fourier transform, to restore the image unaffected by the atmospheric turbulence. In addition, astrometric and photometric algorithms, e.g., StarFinder \citep{diolaiti00} or DAOPHOT \citep{stetson87}, usually also need an estimation of the PSF.

Wavefront-related data delivered by the AO real-time computer can help to accurately estimate the PSF. \citet{veran97} have been the first to develop such a PSF reconstruction algorithm. Implemented on the CFHT curvature sensing AO system PUEO \citep{arsenault94}, it has been routinely delivering reconstructed on-axis PSFs for more than 8 years now. A first attempt to transpose this algorithm to the Shack-Hartmann (SH) wavefront sensor has been undertaken by \citet{harder00}.

Based on the \citet{veran97} algorithm, PSF reconstruction has been developed for three AO systems, equipped this time with SH wavefront sensors, and tested during a few runs of observations, leading to good results

\begin{itemize}
\item \citet{weiss03} has written a piece of PSF reconstruction software for ALFA \citep{kasper00},  the SH AO system of the Calar Alto 3.5m telescope; 
\item \citet{jolissaint04} has written OPERA, a piece of PSF reconstruction software for Altair \citep{herriot00}, the 4-quadrant SH AO system of the Gemini North telescope; and
\item \citet{fitzgerald04} have written a piece of PSF reconstruction software for  the SH AO system of the UCO/Lick ObservatoryÕs 3 m Shane Telescope \citep{bauman02}.
\end{itemize}

These current existing algorithms use particular functions, usually named $U_{ij}$, computed from the mirror modes. The number of $U_{ij}$ functions is proportional to the square number of mirror modes, which leads to gigabytes of data to handle for systems with 150 to 200 actuators, and thus limits the efficiency of the PSF reconstruction process. 

The goal of this article is to propose two new algorithms that avoid the use of these $U_{ij}$ functions. The global scheme of the PSF reconstruction is not affected; we have simply replaced the steps involving the $U_{ij}$ functions. We also provide, as a by-product, a way to estimate the PSF likelihood. The latter may be crucial for image deconvolution algorithms, which currently lack this kind of information.

We present the main points of the PSF reconstruction algorithm developed by \citet{veran97}, as well as their different assumptions in Sect.~\ref{sect:veran}. In Sect.~\ref{sect:new_algo}, we describe the two algorithms we propose, and in Sect.~\ref{sect:test}, the tests from AO simulations. We discuss the choice and use of each algorithm in the last section.


\section{The current $U_{ij}$ algorithm}
\label{sect:veran}
\subsection{The long-exposure AO-corrected PSF expression}
\label{section21}
In the PSF reconstruction algorithm developed by \citet{veran97}, assuming that the phase structure function defined in Eq.~\ref{eq4} below is stationary over the pupil, the AO-corrected monochromatic long-exposure OTF is decomposed as follows:

\begin{equation}
\Big\langle OTF\big(\vec{\rho}/\lambda\big)\Big\rangle =  \Big\langle OTF_{\phi_\epsilon}\big(\vec{\rho}/\lambda\big)\Big\rangle\times OTF_{\mathrm{tel}}\big(\vec{\rho}/\lambda\big)
\label{eq1},
\end{equation}

\noindent where: \begin{itemize}
\item $\Big\langle OTF_{\phi_\epsilon}\big(\vec{\rho}/\lambda\big)\Big\rangle$ is the attenuation of the long-exposure OTF due to the partial correction of AO, and
\item $OTF_{\mathrm{tel}}\big(\vec{\rho}/\lambda\big)$ is the perfect telescope OTF.
\end{itemize}

The phase $\phi_\epsilon$ can be split into two parts: $\phi_{\epsilon_\|}$, which belongs to the vector space spanned by the mirror modes, and $\phi_{\epsilon_\perp}$, which is orthogonal to the former space.

\begin{eqnarray}
 \Big\langle OTF\big(\vec{\rho}/\lambda\big)\Big\rangle=\Big\langle OTF_{\phi_{\epsilon_\|}}\big(\vec{\rho}/\lambda\big)\Big\rangle\times\Big\langle OTF_{\phi_{\epsilon_\perp}}\big(\vec{\rho}/\lambda\big)\Big\rangle\qquad\quad \nonumber\\
{}\times OTF_{\mathrm{tel}}\big(\vec{\rho}/\lambda\big)
\label{eq2}
\end{eqnarray}

and, this expression can be written

\begin{eqnarray}
 \Big\langle OTF\big(\vec{\rho}/\lambda\big)\Big\rangle \approx \exp\big(-\frac{1}{2} \bar{D}_{\phi_{\epsilon_\|}}(\vec{\rho})\big)\times\exp\big(-\frac{1}{2} \bar{D}_{\phi_{\epsilon_\perp}}(\vec{\rho})\big)\quad\nonumber\\
 {}\times OTF_{\mathrm{tel}}\big(\vec{\rho}/\lambda\big)
\label{eq3},
\end{eqnarray}

\noindent where: \begin{itemize}
\item $\Big\langle OTF_{\phi_{\epsilon_\|}}\big(\vec{\rho}/\lambda\big)\Big\rangle$ is the attenuation of the  long-exposure OTF due the mirror component of the phase, i.e., the "residual parallel phase",
\item $\Big\langle OTF_{\phi_{\epsilon_\perp}}\big(\vec{\rho}/\lambda\big)\Big\rangle$ is the attenuation of the  long-exposure OTF due the component of the phase belonging to the space perpendicular to the mirror space, i.e., the "perpendicular phase", 
\item $\bar{D}_{\phi_{\epsilon_\|}}(\vec{\rho})$ is the mean structure function of the residual parallel phase, 
\item $\bar{D}_{\phi_{\epsilon_\perp}}(\vec{\rho})$ is the mean structure function of the perpendicular phase,
\item $\vec{\rho}$ is a pupil plane coordinate vector, and
\item $\lambda$ is the wavelength of observation.
\end{itemize}

This decomposition is based on several assumptions (cf. \citealt{veran97}):

\begin{enumerate}
\item the complex field amplitude is uniform over the pupil (scintillation is neglected), 
\item the residual parallel phase $\phi_{\epsilon_\|}(\vec{x})$ and the perpendicular phase $\phi_{\epsilon_\perp}(\vec{x})$ follow Gaussian statistics,
\item the phase structure function is stationary over the telescope pupil, and
\item the correlation between the mirror component and the perpendicular component of the phase is negligible.
\end{enumerate}

From the expression of the structure function of the residual parallel phase:

\begin{equation}
D_{\phi_{\epsilon_\|}}(\vec{x},\vec{\rho})=\Big\langle\big(\phi_{\epsilon_\|}(\vec{x})-\phi_{\epsilon_\|}(\vec{x}+\vec{\rho})\big)^2\Big\rangle
\label{eq4}
\end{equation}

\noindent and the decomposition of the phase on the basis of the mirror modes $\{M_i(\vec{x})\}_{i=1...N}$: 

\begin{equation}
\phi_{\epsilon_\|}(\vec{x},t)=\sum_{i=1}^N \epsilon_{\|i}(t)\,M_i(\vec{x})
\label{eq5},
\end{equation}

\noindent one obtains:

\begin{equation}
D_{\phi_{\epsilon_\|}}(\vec{x},\vec{\rho})=\sum_{i=1}^N\sum_{j=1}^N \langle\epsilon_{\|i}\epsilon_{\|j}\rangle\big(M_i(\vec{x})-M_i(\vec{x}+\vec{\rho})\big)\big(M_j(\vec{x})-M_j(\vec{x}+\vec{\rho})\big)
\label{eq6}.
\end{equation}

The mean structure function of the residual parallel phase $\bar{D}_{\phi_{\epsilon_\|}}(\vec{\rho})$ is the mean of $D_{\phi_{\epsilon_\|}}(\vec{x},\vec{\rho})$ over $\vec{x}$:

\begin{equation}
\bar{D}_{\phi_{\epsilon_\|}}(\vec{\rho})=\frac{\displaystyle{\int D_{\phi_{\epsilon_\|}}(\vec{x},\vec{\rho})P(\vec{x}) P(\vec{x}+\vec{\rho})\mathrm{d}\vec{x}}}{\displaystyle{\int P(\vec{x})\, P(\vec{x}+\vec{\rho})\mathrm{d}\vec{x}}}
\label{eq7}.
\end{equation}

The expression of $\bar{D}_{\phi_{\epsilon_\perp}}(\vec{\rho})$ is similarly derived from $D_{\phi_{\epsilon_\perp}}(\vec{x},\vec{\rho})$.

The corresponding  AO-corrected monochromatic long exposure PSF is derived as the Fourier transform of the OTF.

\subsection{The $U_{ij}(\vec{\rho})$ functions}

Equation~\ref{eq6} can be rewritten:

\begin{equation}
\bar{D}_{\phi_{\epsilon_\|}}(\vec{\rho})=\sum_{i=1}^N\sum_{j=1}^N \langle\epsilon_{\|i}\epsilon_{\|j}\rangle\,U_{ij}(\vec{\rho})
\label{eq8}
\end{equation}

The $U_{ij}(\vec{\rho})$ functions  are defined by:

\begin{equation}
\frac{\displaystyle{\int \!\big(M_i(\vec{x})-M_i(\vec{x}+\vec{\rho})\big)\big(M_j(\vec{x})-M_j(\vec{x}+\vec{\rho})\big)P(\vec{x}) P(\vec{x}+\vec{\rho})\mathrm{d}\vec{x}}}{\displaystyle{\int P(\vec{x})\, P(\vec{x}+\vec{\rho})\mathrm{d}\vec{x}}}
\label{eq9},
\end{equation}

\noindent where $P(\vec{r})$ is the pupil function and $\vec{x}$ a coordinate vector in the pupil plane.

In practice, using the Fourier transform and the properties of the correlation function \citep{veran97}, the $U_{ij}(\vec{\rho})$ functions are computed as:

\begin{equation}
U_{ij}(\vec{\rho})=\frac{\displaystyle\mathcal{F}^{-1}\bigg(2  \Re\left(\mathcal{F}(M_i M_j P)\mathcal{F}^*(P)-\mathcal{F}(M_i P)\mathcal{F}(M_j P)\right) \bigg)}{\displaystyle\mathcal{F}^{-1}\left(\mid\mathcal{F}(P)\mid^2\right)}
\label{eq9b},
\end{equation}

\noindent where $\mathcal{F}$ is the Fourier transform function and $ \Re$ the complex number real part function.
 
Equation \ref{eq8} is a key one for the experimental reconstruction of PSFs. The covariance matrix $\langle\epsilon_\|{\epsilon_\|}^t\rangle$ has to be measured experimentally on the AO system itself, by averaging the cross-products of wavefront measurements obtained during the time of the image exposure.  

In the current PSF reconstruction algorithms, derived from \citet{veran97}, the matrix $\langle\epsilon_\|{\epsilon_\|}^t\rangle$ is the basic entry point from which one can deduce successively the phase structure function, the OTF, and then the PSF. Additionally, one has to compute, store once for all, and also read the $U_{ij}(\vec{\rho})$ functions during the reconstruction process .

In Eq.~\ref{eq8}, the $i$ and $j$ indices play a symmetric role, so that there are actually $N\times(N+1)/2$ "useful" $U_{ij}(\vec{\rho})$ functions. As an example, in the case of the VLT AO system NAOS \citep{rousset00}, the 159 compensated modes lead to 12720 "useful" $U_{ij}(\vec{\rho})$ functions. Today, the large number of $U_{ij}(\vec{\rho})$ hence represents, depending on the array size and data type, several gigabytes of data to compute, store, and read. Even if in practice, Eq.~\ref{eq8} can be efficiently implemented by Fourier transform (cf. Eq.~\ref{eq9b}), this leads to a heavy PSF reconstruction process, which will turn out to be impossible to handle in the future since the next AO systems are expected to have a largely increased number of modes: about 1370 actuators for the VLT-Planet Finder AO system \citep{fusco05} and  several tens of thousands for extremely large telescopes.  In the following, we propose  a way to achieve this computation, starting from the same covariance matrix $\langle\epsilon_\|{\epsilon_\|}^t\rangle$, without using the $U_{ij}(\vec{\rho})$.

\section{Theory of the proposed algorithms}
\label{sect:new_algo}
\subsection{The $V_{ii}$ algorithm}

Let us consider the vector $\epsilon_\|(t)$, hereafter $\epsilon_\|$, made of the $\{\epsilon_{\|i}\}_{i=1...N}$ coefficients, i.e., the vector representing $\phi_{\epsilon_\|}(\vec{x},t)$ in the basis of the mirror modes $M_i(\vec{x})$. The eigen decomposition of the residual parallel phase covariance matrix is:

\begin{equation}
\Lambda=B^t\langle\epsilon_\|{\epsilon_\|}^t\rangle B
\label{eq10},
\end{equation}

\noindent where $\Lambda$ is a diagonal matrix that contains the $\{\lambda_i\}_{i=1...N}$ eigenvalues and B is the matrix of eigenvectors: $B^tB=BB^t=Id$.

Equation~\ref{eq10} can be written:

\begin{equation}
\Lambda=\big\langle(B^t\epsilon_\|)(B^t\epsilon_\|)^t\big\rangle
\label{eq11}.
\end{equation}

The vector $\eta$ equal to  $B^t\epsilon_\|$ represents $\phi_{\epsilon_\|}(\vec{x},t)$ in the basis that diagonalizes the residual parallel phase covariance matrix. Its coefficients are noted $\{\eta_i\}_{i=1...N}$. From Eq.~\ref{eq11}, the covariance matrix $\langle\eta \eta^t\rangle$ is diagonal, i.e., in this new basis, the residual parallel phase covariance matrix is diagonal, and the mean residual parallel phase structure function reduces to:

\begin{equation}
\bar{D}_{\phi_{\epsilon_\|}}(\vec{\rho})=\sum_{i=1}^N \langle\eta_i\eta_i\rangle\,V_{ii}(\vec{\rho})=\sum_{i=1}^N \lambda_i\,V_{ii}(\vec{\rho})
\label{eq12},
\end{equation}

\noindent where the $V_{ij}(\vec{\rho})$ functions are the equivalent in the new basis to the $U_{ij}(\vec{\rho})$ functions (Eq.~\ref{eq8}). Similarly to Eq.~\ref{eq9}, the $V_{ij}(\vec{\rho})$ functions are defined by

\begin{equation}
\frac{\displaystyle{\int \!\big(M^\prime_i(\vec{x})-M^\prime_i(\vec{x}+\vec{\rho})\big)\big(M^\prime_j(\vec{x})-M^\prime_j(\vec{x}+\vec{\rho})\big)P(\vec{x}) P(\vec{x}+\vec{\rho})\mathrm{d}\vec{x}}}{\displaystyle{\int P(\vec{x})\, P(\vec{x}+\vec{\rho})\mathrm{d}\vec{x}}}
\label{eq12b},
\end{equation}

\noindent such that $\mathcal{M}^\prime$, the matrix made of the eigenvector modes $\{M^\prime_i(\vec{x})\}_{i=1...N}$ is given by $\mathcal{M}^\prime=B^t\mathcal{M}$, $\mathcal{M}$ being the matrix made of the mirror modes $\{M_i(\vec{x})\}_{i=1...N}$.

Using the $V_{ii}$ algorithm, the computation of the residual parallel phase OTF only requires the computation of a number $N$ of functions $V_{ii}(\vec{\rho})$. However, these $V_{ii}(\vec{\rho})$ functions have to be computed on the fly for each estimation of the mean residual parallel phase structure function.

\subsection{The "instantaneous PSF" algorithm}

The solution we propose here is similar to the algorithm presented by \citet{roddier90} to simulate atmospherically distorted wavefronts, calculated from the covariance matrix of the coefficients of their expansion in Zernike modes. We extend this algorithm to any modal basis and covariance matrix, and use it to reconstruct AO-corrected PSFs.

Let us consider again the eigen decomposition of the residual parallel phase covariance matrix:

\begin{equation}
\langle\epsilon_\|{\epsilon_\|}^t\rangle=B\Lambda B^t
\label{eq13},
\end{equation}

If one generates a vector $\eta$ whose coefficients are independent Gaussian random variables with zero mean and variance equal to the eigenvalue $\lambda_i$, i.e., $\langle \eta \eta^t\rangle=\Lambda$,
then the vector $\beta=B\eta$ is a set of correlated random variables whose covariance matrix is  $\langle\epsilon_\|{\epsilon_\|}^t\rangle$: 

\begin{equation}
\langle \beta\beta^t\rangle=\langle B\eta\eta^tB^t\rangle= B\Lambda B^t=\langle\epsilon_\|{\epsilon_\|}^t\rangle
\label{eq14}.
\end{equation}

The phase represented by the vector $ \beta$ is:

\begin{equation}
\phi(\vec{x},t)=\sum_{i=1}^N \beta_i(t)\,M_i(\vec{x}),
\end{equation}

\noindent and the "instantaneous" PSF corresponding to that phase is:
\begin{equation}
PSF_\|(\vec{x},t)=\Big\|\mathcal{F}\big(\exp(i\phi(\vec{x},t)\big)\Big\|^2.
\end{equation}

Then, by generating random $\eta$ vectors such that $\langle \eta \eta^t\rangle=\Lambda$, we build instantaneous PSFs that, on average, converge to the long-exposure PSF of the mirror space. Note that the latter is not the "full PSF" that would be observed at the telescope since it does not include the uncorrected part of the phase (cf. Eq.~\ref{eq2}). Finally:

\begin{equation}
\Big\langle OTF_{\phi_{\epsilon_\|}}\big(\vec{x}/\lambda\big)\Big\rangle\times OTF_{\mathrm{tel}}\big(\vec{x}/\lambda\big)=\mathcal{F}\left(\sum_t PSF_\|(\vec{x},t)\right).
\end{equation}

\section{Test of the algorithms}
\label{sect:test}
\subsection{Description of the simulation}
   \begin{figure}[t]
   \centering
    \resizebox{\hsize}{!}{\includegraphics{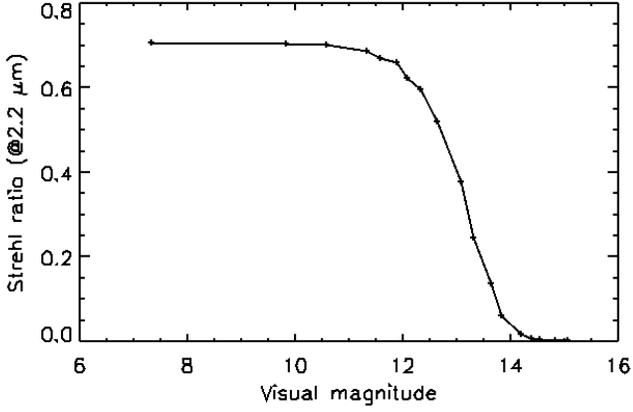}}
      \caption{Strehl ratio at 2.2 $\mu$m vs. guide star visual magnitude for a NAOS-like AO system.}
         \label{fig:strehl}
   \end{figure}
   
To test our algorithms, we have used a Monte Carlo-based AO simulation software
developed at ONERA. It is a complete end-to-end AO simulator divided into four main modules (calibration, propagation, closed-loop, and focal plane imaging modules) that make it as close as possible to an actual system. The complete algorithm is fully described in \citet{conan04} and has been used extensively for the design of AO systems, especially the future
Planet Finder System for the VLT \citep{fusco05}, as well as for the tests and validations of
existing AO systems (NAOS for instance).

We have tested our algorithms in the simple case of  a NAOS-like AO system
\citep{rousset00}: a 14$\times$14 subpupil Shack-Hartmann wavefront sensor
with 8$\times$8 pixels per sub-aperture, a read-out noise of 3 e$^-$ per pixel, and a
sampling frequency of  500 Hz, 2048 loop iterations. The correction was performed with a 185 actuator piezostack deformable mirror plus a tip-tilt mirror. The wavefront sensing and observation wavelengths of the simulation were 0.65 and 2.2 $\mu$m, respectively. The seeing was 0.85\arcsec\ at 0.5 $\mu$m. 

Since we aimed at testing our algorithms with different conditions of
correction, we ran the simulation with a guide star magnitude ranging
from 7 to 15 so that the resulting Strehl ratio ranged from $\sim$70\%
down to $\sim$0.2\% (Fig.~\ref{fig:strehl}). For a given guide star magnitude, we stored all the values  $\epsilon_\|(t)$ obtained from the simulation to compute the  covariance matrix $\langle\epsilon_\|{\epsilon_\|}^t\rangle$ and ran the $U_{ij}$, $V_{ii}$,  and "instantaneous PSF" algorithms (code detailed in Appendix~\ref{appendix:codes}) from this covariance matrix to derive the corresponding OTFs for comparison.

 \subsection{Results of the simulations}
\begin{figure}[t]
   \centering
    \resizebox{\hsize}{!}{\includegraphics{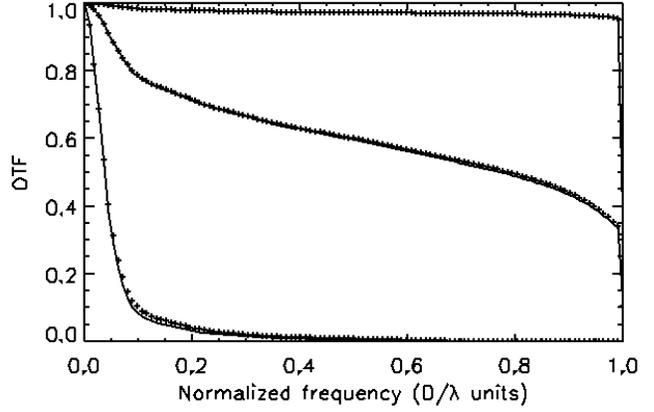}}
      \caption{Circular mean of the "atmospheric" OTFs for different conditions of correction: from top to bottom, the guide star magnitude is 7.3, 12.3, and 13.6. For each guide star magnitude, the solid line is obtained with the $U_{ij}$/$V_{ii}$ and the crosses with the "instantaneous PSF" algorithm".}
         \label{fig:otfcircmoy}
   \end{figure}

\begin{figure}[t]
   \centering
    \resizebox{\hsize}{!}{\includegraphics{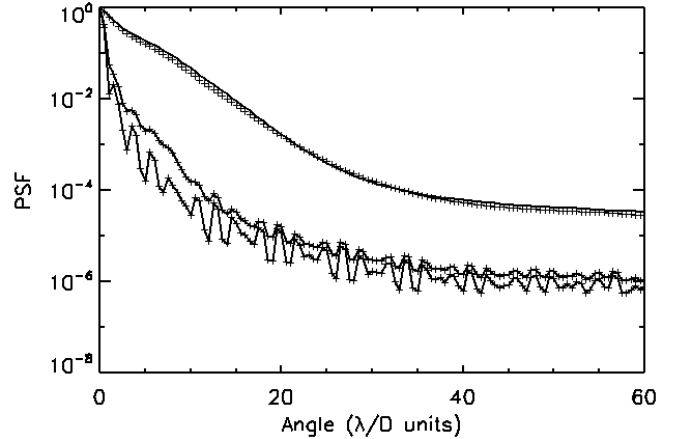}}
      \caption{Circular mean of the PSFs for different conditions of correction: from top to bottom, the guide star magnitude is 13.6, 12.3, and 7.3. For each guide star magnitude, the solid line is obtained with the $U_{ij}$/$V_{ii}$ and the crosses with the "instantaneous PSF" algorithm". Note that the PSFs considered here only correspond to the residuals and do not include the perpendicular part of the phase.}
         \label{fig:psfcircmoy}
   \end{figure}

Figure~\ref{fig:otfcircmoy} represents the circular mean of the residual "atmospheric" OTFs (i.e., not multiplied by the telescope OTF) produced by the different algorithms in three cases: good, moderate and poor correction (7.3$^{th}$, 12.3$^{th}$, and 13.6$^{th}$ magnitude guide stars, respectively). Within the numerical uncertainties, the $U_{ij}$ and $V_{ii}$ algorithms produce exactly the same OTFs, which is not unexpected since they mathematically do the same computations, but in a different modal basis. Their corresponding PSF profiles in Fig.~\ref{fig:psfcircmoy} are consequently also the same.

This is of course not the case for the "instantaneous PSF" algorithm, which needs a large enough number of iterations to converge. However, even in the poor correction case, the OTF profile obtained with the "instantaneous PSF" algorithm well reproduces the $U_{ij}$ profile (Fig.~\ref{fig:otfcircmoy}) with a maximum error of a few 10$^{-2}$ around 0.05 $D/\lambda$. This is also confirmed with the PSF profiles that are almost superimposed, even in the poor correction case (Fig.~\ref{fig:psfcircmoy}).

\section{Discussion}
The $U_{ij}$ and $V_{ii}$ algorithms mathematically produce the same OTFs. In practice, the $U_{ij}$ algorithm requires reading each of the $N\times(N+1)/2$  $U_{ij}$ functions (where $N$ is the number of modes), which have been computed and stored previously, once. In comparison, the $V_{ii}$ algorithm requires to diagonalize the covariance matrix, to compute the new modes $\{M^\prime_i(\vec{x})\}_{i=1...N}$, and to compute each $V_{ii}$ function. This latter step is composed of calls to these functions (cf. Appendix A):
\begin{itemize}
\item 2 FFTs,
\item 1 square function,
\item 1 real part of a complex number, and
\item 1 modulus of a complex number.
\end{itemize}

As an example, the $V_{ii}$ algorithm took $\sim$8 seconds in our simulation: $\sim$2 seconds to diagonalize the covariance matrix and compute the modes (the former taking a negligible time), and $\sim$6 seconds to compute the $N$ $V_{ii}$ functions. This is $\sim$25 times faster than the $U_{ij}$ algorithm ($\sim$191 seconds). In addition to this huge gain in computation time, a large amount of disk space is saved. This causes the $V_{ii}$ algorithm to always be preferred to the $U_{ij}$ one.

As noticed by \citet{conan94}, averaging short-exposure OTFs, as we do in the "instantaneous PSF" algorithm, is a process that converges very slowly, especially at large $D/r_0$ or a low correction level. In addition, it does not lead to the infinitely long exposure OTF since a given number of short exposures are averaged. Besides, \citet{conan94} has shown that in the poor correction case, the error in computing the long-exposure OTF of such an algorithm is larger than for the $U_{ij}$ algorithm, and then the $V_{ii}$ algorithm as well.

Though, we emphasize that, in addition to the OTF computation itself, the "instantaneous PSF" algorithm can provide an estimation about the variability of the OTF,  which can help a lot in some deconvolution algorithms. The estimation of the infinitely long exposure OTF that results from the convergence of our "instantaneous PSF" algorithm and that corresponds to a given covariance matrix $\langle\epsilon_\|{\epsilon_\|}^t\rangle$ is unique: let us call it $OTF_{\infty}(\vec{\rho}/\lambda)$. The dispersion in the random generation of OTFs can be computed as 

\begin{equation}
\sigma^2(\vec{\rho}/\lambda) = \Big\langle \big\|OTF_{\infty}(\vec{\rho}/\lambda) - OTF_{i}(\vec{\rho}/\lambda) \big\|^2 \Big\rangle_{i},
\end{equation} 
where $OTF_i$ is the $i^{th}$ draw of a randomly-generated OTF.
If we call $OTF_\mathrm{{obs}}(\vec{\rho}/\lambda)$ the OTF actually observed on the instrument during the given, non infinite, observing time $T_\mathrm{{int}}$, when the given covariance matrix $\langle\epsilon_\|{\epsilon_\|}^t\rangle$ was measured, we can evaluate how far our estimation $OTF_{\infty}$ is from $OTF_\mathrm{{obs}}$ by writing:

\begin{equation}
\Big\langle \big\|OTF_{\infty}(\vec{\rho}/\lambda) - OTF_\mathrm{{obs}}(\vec{\rho}/\lambda) \big\|^2 \Big\rangle = 
\frac{\sigma^2(\vec{\rho}/\lambda)}{n},
\end{equation}
where $n$ is the equivalent number of independent realisations of PSFs, whose sum has resulted in the final PSF observed by the instrument during the given, non infinite integration time $T_\mathrm{{int}}$. An estimation of $n$ could be obtained for example from a full simulation of the AO system under the same atmospheric conditions as during the observation. It is also be reasonable to consider that the impact of the correction by the AO system is to shorten the image correlation time compared to the image correlation time $\tau_0(\lambda)$ of the atmospheric seeing \citep{rigaut91}. We can then find a lower bound given by $n> T_\mathrm{{int}}/\tau_0(\lambda)$.

\appendix
\section{IDL codes}
\label{appendix:codes}
\subsection{Common parts}

\hspace{4mm} cbmes = readfits('cbmes.fits')

s2m = readfits('slopes2modes.fits')
 
modes = readfits('modalbasis.fits')

nmodes=(size(modes))(3)

lambda\_im=2.2d-6

lambda\_aso=0.65d-6

ratio\_lambda=lambda\_aso/lambda\_im

\noindent ; Telescope OTF computation and corresponding mask

apert=readfits('pupille.fits')

dim=(size(apert))(1)

dim2=2*dim

pup=dblarr(dim2,dim2)

pup(0,0)=apert

pupfft=fft(double(pup))

conjpupfft=conj(pupfft)

otftel = real\_part(fft(pupfft*conjpupfft,/inverse))

den=1/(otftel)

den(where(finite(den) eq 0))=0

mask=fltarr(dim2,dim2)+1

mask(where(otftel lt 1e-5))=0

otftel=otftel/max(otftel)

\noindent ;Computation of the covariance matrix over the modes

cbmode = cbmes\#\#transpose(s2m)

nbcb = (size(cbmode))(2)

covmode = (cbmode\#transpose(cbmode))/nbcb

\subsection{The $U_{ij}$ algorithm}

\noindent ; Computation and storage of the Uij functions

modei = dblarr(dim2,dim2)

modej = dblarr(dim2,dim2)

for i=0,nmodes-1 do begin

    	\hspace{4mm} modei(0,0) = modes(*,*,i)*apert 

    	\hspace{4mm} for j=i,nmodes-1 do begin

       		\hspace{8mm} modej(0,0) = modes(*,*,j)*apert

       		\hspace{8mm} term1 = real\_part(fft(modei*modej)*conjpupfft)

       		\hspace{8mm} term2 = real\_part(fft(modei)*conj(fft(modej)))

       		\hspace{8mm} \mbox{uij = real\_part(fft(2*(term1-term2),/inverse))*den*mask}

		\hspace{8mm} writefits,'Uij/u\_'+string(i,format='(i2.2)')+'\_'+\$

		\hspace{30mm} string(j,format='(i2.2)')+'.fits',uij

    	\hspace{4mm} endfor

endfor

\noindent ; Computation of the OTF with the Uij

dph1 = dblarr(dim2,dim2)

for i=0,nmodes-1 do begin

\hspace{4mm} for j=i,nmodes-1 do begin

\hspace{8mm} uij = readfits('Uij/u\_'+string(i,format='(i2.2)')+'\_'+\$

\hspace{20mm} string(j,format='(i2.2)')+'.fits',/silent)

\hspace{8mm} fac = double((i ne j)+1)

\hspace{8mm} dph1 = dph1+(fac*covmode(i,j)*uij)

 \hspace{4mm}  endfor

	endfor

	otf1 = exp(-0.5*dph1*ratio\_lambda\verb+^+2)*mask 

	otf1 = otf1/max(otf1)

\vspace{-2mm}

\subsection{The $V_{ii}$ algorithm}
; New modes that diagonalize the covariance matrix

l = (eigenql(covmode,eigenvectors=b))$>$0

s = b\#diag\_matrix(sqrt(l))

newmodes = reform((reform(modes,dim\verb+^+2,nmodes))\#b,\$

\hspace{55mm} dim,dim,nmodes)

\noindent ; Computation of the OTF with the Vii

newmodei = dblarr(dim2,dim2)

temp = dblarr(dim2,dim2)

for i=0,nmodes-1 do begin 

\hspace{4mm} 	newmodei(0,0) = newmodes(*,*,i)*apert

\hspace{4mm}        	term1 = real\_part(fft(newmodei\verb+^+2)*conjpupfft) 

\hspace{4mm}        	term2 = (abs(fft(newmodei)))\verb+^+2 

\hspace{4mm}        	temp = temp+((term1-term2)*l(i)) 

endfor 

otf2 = exp(-0.5*real\_part(fft(2*temp,/inverse))*den\$

\hspace{40mm} *ratio\_lambda\verb+^+2)*mask 

otf2 = otf2/max(otf2) 

\vspace{-1mm}

\subsection{The "instantaneous PSF" algorithm}

\hspace{4mm}		psf = dblarr(dim2,dim2)

	tmp = ratio\_lambda*reform(modes,dim\verb+^+2,nmodes)
	
	for i=0,nbcb-1 do begin
	
\hspace{4mm}		phi(0,0) = reform((s\#randomn(seed,nmodes))\#\#tmp,dim,dim)
	
\hspace{4mm}		psf = psf+(abs(fft(pup*exp(dcomplex(0,phi)))))\verb+^+2 
	
	endfor

	psf = psf/nbcb
	
	otf3 = real\_part(fft(psf))
	
	otf3 = otf3/max(otf3)/otftel*mask 
	
	otf3(where(finite(otf3) eq 0))=0

\end{document}